\documentclass[aps,prc,twocolumn,superscriptaddress]{revtex4}

\usepackage{graphicx}% Include figure files
\usepackage{dcolumn}% Align table columns on decimal point
\usepackage{bm}% bold math

\begin{document}

% Use the \preprint command to place your local institutional report
% number in the upper righthand corner of the title page in preprint mode.
% Multiple \preprint commands are allowed.
% Use the 'preprintnumbers' class option to override journal defaults
% to display numbers if necessary
%\preprint{}

%Title of paper
\title{Differential cross section of the charge-exchange reaction $\pi^- p \to \pi^0 n$ in the momentum range from~148~to~323~MeV/$c$}

% repeat the \author .. \affiliation  etc. as needed
% \email, \thanks, \homepage, \altaffiliation all apply to the current
% author. Explanatory text should go in the []'s, actual e-mail
% address or url should go in the {}'s for \email and \homepage.
% Please use the appropriate macro foreach each type of information

% \affiliation command applies to all authors since the last
% \affiliation command. The \affiliation command should follow the
% other information
% \affiliation can be followed by \email, \homepage, \thanks as well.
%\author{}
%\email[]{Your e-mail address}
%\homepage[]{Your web page}
%\thanks{}
%\altaffiliation{}
%\affiliation{}

\author{M.~E.~Sadler}
\affiliation{Abilene Christian University, Abilene, TX 79699-7963}

\author{A.~Kulbardis}
\affiliation{Petersburg Nuclear Physics Institute, Gatchina, Russia 188350}

\author{V.~Abaev}
\affiliation{Petersburg Nuclear Physics Institute, Gatchina, Russia 188350}

\author{C.~Allgower}
\affiliation{Argonne National Laboratory, Argonne, IL 60439-4815}

\author{A.~Barker}
\affiliation{Abilene Christian University, Abilene, TX 79699-7963}

\author{V.~Bekrenev}
\affiliation{Petersburg Nuclear Physics Institute, Gatchina, Russia 188350}

\author{C.~Bircher}
\affiliation{Abilene Christian University, Abilene, TX 79699-7963}

\author{W.~J.~Briscoe}
\affiliation{The George Washington University, Washington, DC 20052-0001}

\author{R.~Cadman}
\affiliation{Argonne National Laboratory, Argonne, IL 60439-4815}

\author{C.~Carter}
\affiliation{Abilene Christian University, Abilene, TX 79699-7963}

\author{M.~Clajus}
\affiliation{University of California, Los Angeles, CA 90095-1547}

\author{J.~R.~Comfort}
\author{K.~Craig}
\affiliation{Arizona State University, Tempe, AZ 85287-1504}

\author{M.~Daugherity}
\affiliation{Abilene Christian University, Abilene, TX 79699-7963}

\author{B.~Draper}
\affiliation{Abilene Christian University, Abilene, TX 79699-7963}

\author{D.~Grosnic}
\affiliation{Valparaiso University, Valparaiso, IN 46383-6493}

\author{S.~Hayden}
\affiliation{Abilene Christian University, Abilene, TX 79699-7963}

\author{J.~Huddleston}
\affiliation{Abilene Christian University, Abilene, TX 79699-7963}

\author{D.~Isenhower}
\affiliation{Abilene Christian University, Abilene, TX 79699-7963}

\author{M.~Jerkins}
\affiliation{Abilene Christian University, Abilene, TX 79699-7963}

\author{M.~Joy}
\affiliation{Abilene Christian University, Abilene, TX 79699-7963}

\author{N.~Knecht}
\affiliation{University of Regina, Saskatchewan, Canada S4S OA2}

\author{D.~D.~Koetke}
\affiliation{Valparaiso University, Valparaiso, IN 46383-6493}

\author{N.~Kozlenko}
\affiliation{Petersburg Nuclear Physics Institute, Gatchina, Russia 188350}

\author{S.~Kruglov}
\affiliation{Petersburg Nuclear Physics Institute, Gatchina, Russia 188350}

\author{T.~Kycia (deceased)}
\affiliation{Brookhaven National Laboratory, Upton, NY 11973}

\author{G.~Lolos}
\affiliation{University of Regina, Saskatchewan, Canada S4S OA2}

\author{I.~Lopatin}
\affiliation{Petersburg Nuclear Physics Institute, Gatchina, Russia 188350}

\author{D.~M.~Manley}
\affiliation{Kent State University, Kent, OH 44242-0001}

\author{R.~Manweiler}
\affiliation{Valparaiso University, Valparaiso, IN 46383-6493}

\author{A.~Marusic}
\author{S.~McDonald}
\author{B.~M.~K.~Nefkens}
\affiliation{University of California, Los Angeles, CA 90095-1547}

\author{J.~Olmsted}
\affiliation{Kent State University, Kent, OH 44242-0001}

\author{Z.~Papandreou}
\affiliation{University of Regina, Saskatchewan, Canada S4S OA2}

\author{D.~Peaslee}
\affiliation{University of Maryland, College Park, MD 20742-4111}

\author{J.~Peterson}
\affiliation{University of Colorado, Boulder, CO 80309-0390}

\author{N.~Phaisangittisakul}
\author{S.~N.~Prakhov}
\author{J.~W.~Price}
\affiliation{University of California, Los Angeles, CA 90095-1547}

\author{A.~Ramirez}
\affiliation{Arizona State University, Tempe, AZ 85287-1504}

\author{C.~Robinson}
\affiliation{Abilene Christian University, Abilene, TX 79699-7963}

\author{A.~Shafi}
\affiliation{The George Washington University, Washington, DC 20052-0001}

\author{H.~Spinka}
\affiliation{Argonne National Laboratory, Argonne, IL 60439-4815}

\author{S.~Stanislaus}
\affiliation{Valparaiso University, Valparaiso, IN 46383-6493}

\author{A.~Starostin}
\affiliation{Petersburg Nuclear Physics Institute, Gatchina, Russia 188350}
\affiliation{University of California, Los Angeles, CA 90095-1547}

\author{H.~M.~Staudenmaier}
\affiliation{Universit\"at Karlsruhe, Karlsruhe, Germany 76128}

\author{I.~Strakovsky}
\affiliation{The George Washington University, Washington, DC 20052-0001}

\author{I.~Supek}
\affiliation{Rudjer Boskovic Institute, Zagreb, Croatia 10000}

\author{W.~B.~Tippens}
\affiliation{University of California, Los Angeles, CA 90095-1547}

\author{S.~Watson}
\affiliation{Abilene Christian University, Abilene, TX 79699-7963}

%Collaboration name if desired (requires use of superscriptaddress
%option in \documentclass). \noaffiliation is required (may also be
%used with the \author command).
%\collaboration can be followed by \email, \homepage, \thanks as well.
\collaboration{Crystal Ball Collaboration}
\noaffiliation

\date{\today}

\begin{abstract}
% insert abstract here
Measured values of the differential cross section for pion-nucleon charge exchange, $\pi^- p \to \pi^0 n$, are presented at $\pi^-$ momenta of 148, 174, 188, 212, 238, 271, 298, and 323 MeV/$c$, a region dominated by the $\Delta$(1232) resonance.  
Complete angular distributions were obtained using the
Crystal Ball detector at the Alternating Gradient Synchrotron (AGS) at Brookhaven National Laboratory (BNL).   Statistical uncertainties of the differential cross sections are typically 2-6\%, exceptions being the results at the lowest momentum and at the most forward measurements at the five lowest
momenta.  We estimate the systematic uncertainties to be 3-6\%. 

\end{abstract}

% insert suggested PACS numbers in braces on next line
% insert suggested keywords - APS authors don't need to do this
%\keywords{}

%\maketitle must follow title, authors, abstract, \pacs, and \keywords
\maketitle

% body of paper here - Use proper section commands
% References should be done using the \cite, \ref, and \label commands
%\section{}
% Put \label in argument of \section for cross-referencing
%\section{\label{}}
%\subsection{}
%\subsubsection{}

\section{Introduction}

Several authors \cite{lit1,lit2,lit3,lit4,lit5,lit6,lit7,lit8} have measured $\pi^- p \to \pi^0 n$ differential cross sections in this momentum range.  The previous data were taken using either neutron counters or $\gamma$-ray spectrometers with small solid angle acceptance.  We are adding 160 new data points for the differential cross section taken with the Crystal Ball multi-photon spectrometer, which almost doubles the database in this momentum interval. The Crystal Ball provides complete angular coverage at these momenta by measuring the energy and impact location of the $\gamma$ rays from $\pi^0$ decay. The detector efficiencies inherent in neutron detection are eliminated and the acceptance corrections associated with small detectors are reduced.

Precise data for pion-nucleon charge exchange (CEX) are of interest principally to obtain an accurate description of the $\pi$N system via a consistent and complete set of scattering amplitudes.  A partial-wave analysis (PWA) is typically used, but potential models and Lagrangians based on chiral perturbation theory are often used at low energy.  These approaches are based on all reliable scattering data in the three channels that are experimentally accessible, $\pi^+ p \to \pi^+ p$ and $\pi^- p \to \pi^- p$ elastic scattering and CEX.  These reactions are described by amplitudes $F_{+}$, $F_{-}$ and $F_{CEX}$, respectively.  Assuming isospin invariance, these amplitudes are related by

\begin{center}

$$F_{CEX} = \frac{1}{\sqrt{2}}(F_{+} - F_{-})$$

\end{center}

Isospin symmetry is broken by electromagnetic effects and the up-down quark mass difference.  Mass differences between the neutron and proton and the charged and neutral pions are manifestations of these effects.  Gibbs, Ai and Kaufman\cite{Gibbs} incorporated these mass differences and Coulomb corrections in a coupled-channel potential model.  They included data up to T$_\pi$ = 50 MeV (P$_\pi$ = 128 MeV/$c$), just below the range of data reported here.  A surprising 7\% breaking of isospin invariance was obtained at 40 MeV (113 MeV/$c$).  Similar isospin breaking was reported by Matsinos\cite{Matsinos} using data up to T$_\pi$ = 100 MeV (P$_\pi$ = 197 MeV/$c$) that overlaps with the data reported here.  Fettes and Meissner\cite{Fettes01,Meissner01} investigated isospin breaking in the framework of chiral perturbation theory up to 100 MeV/$c$ and obtained only a 0.7\% effect in the s-waves, where Ref.~\cite{Gibbs,Matsinos} observed the largest effect.  In all three analyses the data for CEX were the most limited in quantity.

The $\Delta^{0}$-$\Delta^{++}$ mass and width differences are of interest to test calculations for isospin-breaking effects in hadrons, particularly the up-down quark mass difference.  The Particle Data Group \cite{pdg} includes three determinations of these differences \cite{Pedroni78,Bernicha96,Abaev95}.  Results from Ref.~\cite{Pedroni78,Bernicha96} are both based on the total cross-section measurements for $\pi^\pm p$ from Pedroni, {\it et al.}\cite{Pedroni78}.  The energy independent partial-wave analysis of Abaev and Kruglov \cite{Abaev95} determined the isospin-$\frac{3}{2}$ phase shifts from $\pi^+ p \to \pi^+ p$ elastic scattering data and again from $\pi^- p \to \pi^- p$ and $\pi^- p \to \pi^0 n$.  Both measurements are needed in the latter case since the $\pi^- p$ reactions involve both isospin $\frac{1}{2}$ and $\frac{3}{2}$.  The uncertainties in this determination were dominated by the existing CEX data.

Another example of the impact of $\pi$N measurements on baryon structure is the
$\pi$N sigma term, which is a measure of chiral symmetry breaking in the strong
interaction.  It is obtained by the extrapolation of the s-wave pion-nucleon
scattering amplitudes to a negative energy point by taking advantage of their
analytic properties.  CEX data affect the determination of the sigma term
indirectly, but are important to provide a stable database to determine the
amplitudes as close to threshold as possible before extrapolating to the
non-physical region. Recent discussions of the sigma term can be found in the
$\pi$N Newsletter\cite{Sainio02,Pavan02,Olsson02,Stahov02}.  Reference \cite{Matsinos} questions the determinations of the low-energy hadronic constants, including the $\pi$N sigma term, in a framework that does not include isospin breaking.

The $\pi$N scattering amplitudes extracted by PWA's provide us with the 
best available information on the $\pi$NN coupling constant and 
the $\pi$N scattering lengths.  
Existing CEX data are more sparse and generally less precise than for $\pi^\pm p \to \pi^\pm p$ elastic scattering.  The data reported here remedy this situation in the $\Delta$ resonance region and below.

\section{Experimental Setup}

\begin{figure}
 \resizebox{0.45\textwidth}{!}{\includegraphics{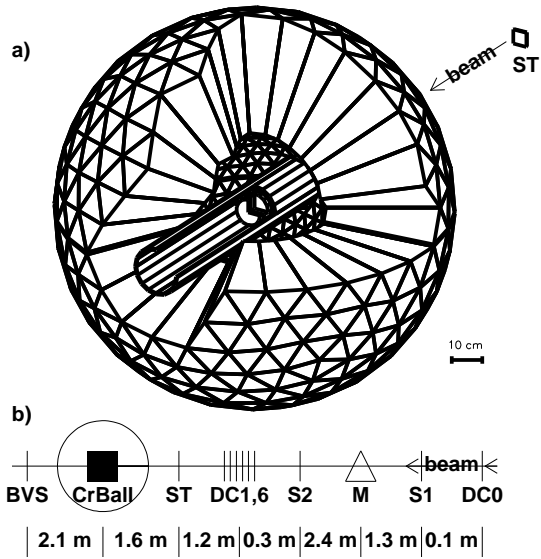}}
\caption{a) The Crystal Ball detector with 1/4 of the crystals in the top hemisphere
removed to show the veto barrel and the target, and b) Schematic picture of the beam line
showing the positions of scintillators S1, ST and BVS and the six downstream drift
chambers (DC1,6). An upstream drift chamber (DC0) was located just before S1.  Between S1 and S2
were a bending magnet (M) and two quadrupoles (not shown).}
 \label{fig:cryball}
\end{figure}

The Crystal Ball (CB) detector (see Fig.~\ref{fig:cryball}) was built by SLAC \cite{slac}
in the 1970's and was used in several experiments at SLAC and DESY.  The CB was
moved to the C6 beam line of the Alternating Gradient Synchrotron (AGS) at Brookhaven National
Laboratory (BNL). The data presented in this work were taken in October, 1998.

The Crystal Ball detector consists of 672 optically isolated NaI(Tl) crystals, a subset
of 720 crystals that would complete a sphere.  The openings for beam entrance and exit
reduce the geometric acceptance to 93\% of $4\pi$ steradians.  The complete sphere is
approximated by an icosahedron consisting of 20 equivalent equilateral major triangles,
each of which is divided into four minor triangles of nine crystals. The individual
crystal dimensions vary slightly depending on their location within a minor triangle.
They are truncated triangular pyramids, nominally 5 cm on edge at the inner radius, 13 cm
at the outer radius, and 41 cm long. Each crystal is viewed by a single photomultiplier
tube. The inner radius of the sphere of crystals is 25 cm.  More detail on the CB is
given in Ref.\cite{Starostin01}.

The cavity in the center of the CB housed a liquid hydrogen (LH$_2$) target.  The target
geometry was a 10-cm diameter cylinder with spherical endcaps. The target length was 10.6
cm along the central beam axis. The target vacuum was maintained inside a cylindrical
aluminum beam pipe (OD = 15.2 cm) with a thickness of 2.1 mm.

ST was the primary beam-defining scintillator and was placed just upstream of the
entrance to the beam pipe. The dimensions were 5.1 cm x 5.1 cm x 0.42 cm (thickness) for
the measurements reported here. It was viewed by two photomultiplier tubes to provide
better timing resolution since all other signals were timed with respect to it.

A veto barrel (VB) was installed to reject events that had charged particles in the final state.
It was constructed of four curved plastic scintillators that formed a cylindrical shell
around the beam pipe. Each segment was 5 mm thick and 120 cm long. Each end of the four
segments was viewed by a photomultiplier tube. The VB logic was formed by the logical AND
of the two ends for a given segment followed by the OR of the four segments.

The {\it Neutral Event} trigger for the experiment was formed by
%\begin{center}
\\
$$Neutral\ Event\ =\rm S1\bullet \rm S2\bullet \rm ST\bullet \overline{\rm WV} \bullet \overline{\rm BH} \bullet\overline{\rm VB}\bullet \rm CB,$$
%\end{center}

where S1, S2 and ST were the beam defining scintillators (see Fig.~\ref{fig:cryball}). WV
and BH are not shown in Fig.~\ref{fig:cryball}.  WV (for wavelength-shifting scintillator) was the
logical OR of four trapezoidal counters that covered the crystals at the entrance tunnel
to the CB in order to veto muons from $\pi^-$ decay.  BH was the logical OR of four beam halo
veto counters that were located around ST.  The purpose of the WV and BH counters was to
prevent accepting triggers from beam particles that hit ST but were within the
(accidental) coincidence time of another beam-associated particle that would deposit
energy in the CB. VB was used to veto charged particles produced in the target,
predominantly from $\pi^-p$ elastic scattering. CB represents the discriminator output of
the fast analog sum of the NaI crystals except the edge crystals surrounding the entrance
and exit tunnels. The discriminator threshold for CB was variable, but corresponded to an
energy of 75 MeV for the data presented here.   A $Charged\ Event$ trigger in which the VB was
put in coincidence was also used.  The number of charged triggers accepted by the data
acquisition was reduced by a factor of
10 using a prescaler. Two beam veto scintillators downstream of
the target, BV and BVS, were not used in the neutral event trigger but the location of
BVS is shown in Fig.~\ref{fig:cryball} because it was used for TOF measurements of the
beam composition (described below).

Pion beam trajectories were measured by the six drift chambers between S2 and ST (three
for the horizontal coordinate and three for the vertical coordinate).  The drift chamber
before the last beam bending magnet determined the difference in momentum of the beam
particle from the nominal value set by the beam tune.  A narrow $\Delta$P/P tune was used
in the experiment.  The width of the momentum distribution was measured
to be 1.4\% (rms), or $\Delta$P/P = 3.4\% (FWHM).

\section{Data analysis}

\subsection{Procedure}

The $\pi^- p \to \pi^0 n$ reaction was identified by measuring the energy and direction of
the two photons from $\pi^0 \to \gamma \gamma$ decay (BR = 98.8\%).  Each photon produces an
electromagnetic shower in the NaI that spreads over several crystals around a central
one. The cluster algorithm finds the crystal with maximum deposited energy and identifies it as the
central one.  A cluster
was defined to be the central crystal and its nearest neighbors.  Clusters with a
central crystal energy greater than 7 MeV and an energy sum over all crystals in the cluster of
at least 17.5 MeV were standard in this analysis.

The direction of the photon is determined by
calculating the trajectory from the target center to the weighted average of the crystal
positions, where the weighting factor is the square root of the deposited energy. The
remaining crystals are searched to find the one with maximum energy to form the next
cluster using the same criteria. The process is repeated until all the clusters are
found.  

With the assumption that the clusters originated from photons at target center, the
invariant mass of photon pairs was found and compared to the $\pi^0$ mass.  Two-cluster
events that had an invariant mass between 97 and 181 MeV/$c^2$ were selected in the
analysis. The recoil neutron can also give a cluster.  Three-cluster events were included
if two of the clusters reconstructed to the $\pi^0$ mass within the same interval and if
the location of the third cluster was consistent with the direction of the neutron. In
principle, this procedure eliminates the need to determine the detection efficiency for
neutrons in the NaI since the events are included in the yield regardless of whether the
neutron is detected.  The efficiency depends strongly on the threshold and increases with
the neutron energy\cite{Stanislaus01}. The percentage of three-cluster events was 1.6\%
at 148 MeV/$c$ and increases to 8.3\% at 298 MeV/$c$. 

The missing mass for producing the two clusters was calculated using the beam momentum
information provided by the drift chambers.  The missing mass was required to be within
110 MeV of the neutron mass.
If this test was passed, the c.m. scattering angle of the $\pi^0$ was calculated
and the data were histogrammed into 20 bins of $\cos \theta_{cm}$.
Runs with an empty target were taken at each momentum and yields were subtracted from
the data taken with the full target.  

The analysis was done in two ways.  The ``full-geometry" analysis included clusters for
which the central crystal was on the edge bordering the entrance or exit tunnels.  The
``near-edge-cut" analysis rejected these events. These analyses required different
calculations of the acceptance, which is discussed in the next section. 

The average pathlength of the pion beam in the LH$_2$ target was calculated using the
trajectories determined from the drift chambers.  All yields were corrected for empty
target normalized to the live-time corrected beam monitor ($S1\bullet \rm S2\bullet \rm ST$).
Since the target was emptied
by displacing the hydrogen liquid with gas, the density of hydrogen gas was subtracted
from the density of liquid (0.0711 g/cm$^3$ at 21 K and 16 psi).  Upon emptying, the temperature of the gas
increased gradually to 60 K so the gas density at 30 K was used, giving a
correction of (1.1 $\pm$ 0.5)\%.

\subsection{Monte Carlo Simulation and Acceptance Calculation}

The acceptance of the Crystal Ball for detecting $\pi^0$'s from $\pi^- p \to \pi^0 n$ was
calculated using a Monte Carlo program based on GEANT \cite{geant}.  All 672 crystals,
the CB enclosure, the target assembly, the beam pipe, and all scintillation counters in
the trigger were included in the simulation. This simulation was used for several
purposes: 1) to calculate the acceptance for $\pi^0$'s in the Crystal Ball for the
different bins in $\cos\theta_{cm}$, 2) to evaluate the fraction of events that
would trigger the veto system, particularly the veto barrel that surrounded the target,
and 3) to gain insight and confidence in the performance of the CB, such as using it to
calibrate the beam momentum as discussed in the next section. 

A separate program (DECKIN) selected a random interaction point in the LH$_2$ target
along the measured beam trajectories that had been saved from the experimental data.
DECKIN then selected outgoing $\pi^0$'s from a given angular distribution and determined energy
and direction of both final-state particles from two-body kinematics.  This information
was passed to the GEANT simulation program (CBall).  The two photons
from $\pi^0 \to \gamma \gamma$ and the neutron were tracked through all elements on which
they were incident and the deposited energy was recorded.  The Monte Carlo events were
then analyzed in the same way as the real data. The average acceptance for a given bin
was the ratio of the number of events that passed the cuts divided by the number thrown.

The two photons and neutron traversed the LH$_2$
target, the containment vessel, beam pipe and veto barrel scintillator before reaching
the Crystal Ball. The photons could convert to $e^+e^-$ or the neutrons could interact
hadronically in any of these materials.  The veto barrel rejected these events if 
the energy deposited exceeded the signal threshold.  This threshold was low in order to reject
minimum ionizing charged particles, so this correction was significant.  It was evaluated
as part of the Monte Carlo simulation.

\begin{figure*}
 \resizebox{0.90\textwidth}{!}{\includegraphics{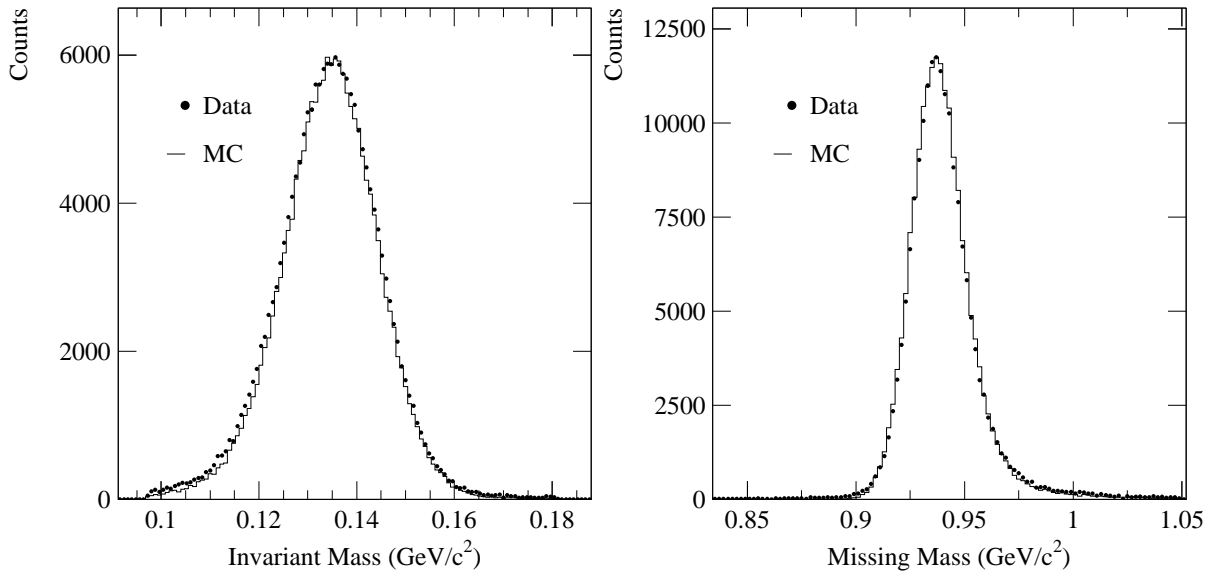}}
\caption{Comparison between data and Monte Carlo of the invariant mass and missing mass
distributions for $\gamma\gamma$ clusters at 298 MeV/$c$.
The normalized target empty subtraction was applied to the data.}
 \label{fig:im_mm_298_ne}
\end{figure*}

\begin{figure*}
 \resizebox{0.90\textwidth}{!}{\includegraphics{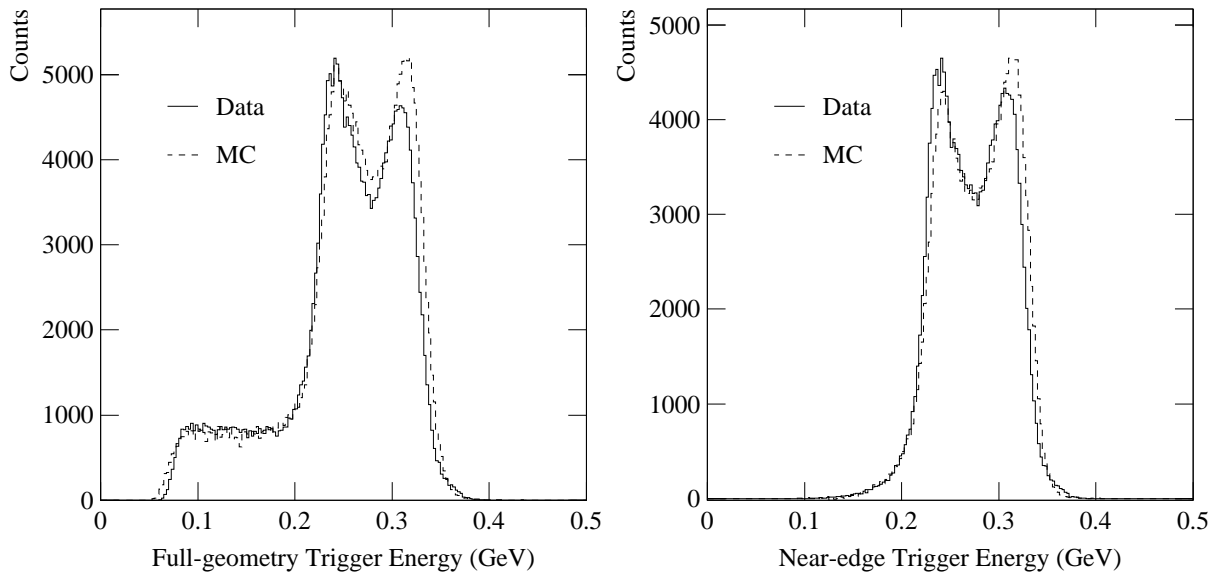}}
\caption{Comparison between data and Monte Carlo of the total trigger energy 
at 298 MeV/$c$ for the full-geometry analysis (left) and the near-edge-cut analysis (right).}
 \label{fig:trig_298}
\end{figure*}

\begin{figure}
 \resizebox{0.45\textwidth}{!}{\includegraphics{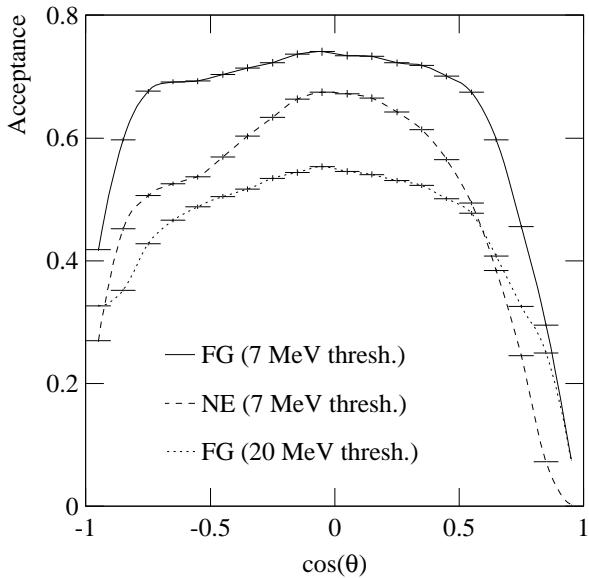}}
\caption{Comparison of the Monte Carlo acceptances at 298 MeV/$c$ for the following
analyses: 1) full geometry (FG) with a 7 MeV crystal threshold, 2) near edge (NE) cut with a 7 MeV
crystal threshold and 3) full geometry with a 20 MeV crystal threshold.}
 \label{fig:Acceptance_298}
\end{figure}

\begin{figure}
 \resizebox{0.45\textwidth}{!}{\includegraphics{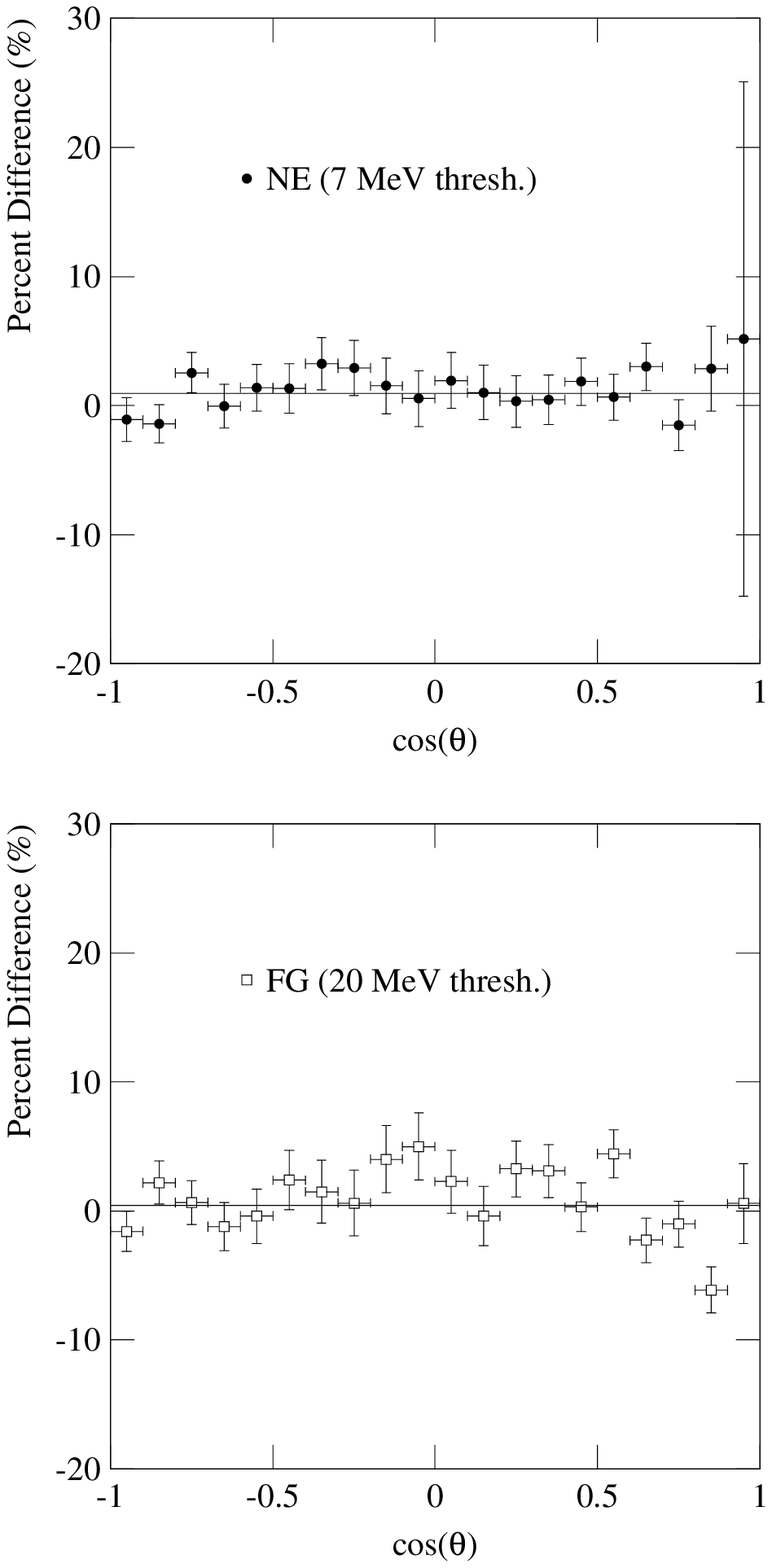}}
\caption{Comparison of the acceptance-corrected yields for the different analyses at 298 MeV/$c$.
Top: Percentage difference between the near-edge-cut (NE) and full geometry analyses, both using 7 MeV crystal
thresholds.  Bottom: Comparison between the two full-geometry (FG) analyses using different crystal
thresholds of 7 and 20 MeV.  The horizontal line indicates the weighted average for each plot and is less
than 1\% for each case.}
 \label{fig:cxs_comp_298}
\end{figure}

Calibration of the veto barrel was accomplished using special runs for $\pi^+ p \to \pi^+ p$
scattering taken with a charged trigger.  The CB was used to determine the direction of the
outgoing $\pi^+$ and the position
at which it traversed the veto barrel.  Comparison to Monte Carlo simulation of the same
events provided the relationship between the energy deposited and the signal pulse heights recorded for all eight photomultiplier tubes.
The attenuation length of the scintillation light along the veto barrel was determined from
the correlation of the pulse height measured at both ends with the position that was determined for the $\pi^+$.  In the simulation for $\pi^- p \to \pi^0 n$, this attenuation was applied to any energy that was deposited in the veto barrel and compared to the signal threshold.  Simulated events that satisfied the VB logic were counted as charged events. 

A gauge of the performance of the CB and of the Monte Carlo simulation is demonstrated in
the invariant mass distribution in Fig.~\ref{fig:im_mm_298_ne}.  The rms width of the
distribution is 12.1 MeV, or 9.0\%.  An energy resolution of 1.74\%/E$^{0.315}$,
where E is the crystal energy in GeV, was applied in the simulation.  Comparison
of the missing mass distribution is also shown in Fig.~\ref{fig:im_mm_298_ne}.

Comparison between data and results of the simulation for the total trigger energy (the energy deposited in all crystals except the
edge crystals) is shown in Fig.~\ref{fig:trig_298}.  The full-geometry analysis was used for the comparison on the left.  The plateau below 0.2 GeV is caused by events in which significant energy was deposited in a guard crystal.  The trigger threshold
at 75 MeV is readily seen on this plot.  The trigger energy for the near-edge-cut
analysis is shown on the right in Fig.~\ref{fig:trig_298}, where the plateau is replaced by a tail in both the data and the simulation.
The two-peak structure between 0.20 and 0.35 GeV in Fig.~\ref{fig:trig_298} reflects the parabolic angular
distribution of the differential cross section at this momentum as shown in the results below.
The cross section peaks at forward angles where the laboratory energy of the $\pi^0$ is highest,
and also at backward angles where this energy is lowest.  The small difference in the relative height of the the peaks is due to the difference between what is used in the simulation and what is measured.  The ability of the simulation to reproduce in detail the invariant mass, missing mass and trigger energy under different conditions give confidence that it can be used to determine the acceptance.

The acceptance as a function of $\cos\theta_{cm}$ is shown for both full-geometry and
near-edge-cut analyses in Fig.~\ref{fig:Acceptance_298} at 298 MeV/$c$.  At this momentum 14\% of the events were rejected due to inclusion of the veto barrel in the simulation.  The acceptance for
the near-edge-cut analysis falls almost to zero for the most forward angle bin at this
momentum.  The full-geometry analysis must be used here for a reasonable measurement of
the differential cross section near $0^\circ$. The full-geometry acceptance using a 20 MeV
crystal threshold is also shown.  This acceptance is 30\% lower than
that using the lower threshold.

The acceptance-corrected yields for the three analyses shown in Fig.~\ref{fig:Acceptance_298}
are compared in Fig.~\ref{fig:cxs_comp_298}.  Using the full-geometry, 7-MeV threshold analysis as the standard, the percent difference of the near-edge-cut analysis is shown in the top half of the figure.  The line is drawn at the average difference, which was just less than 1\%.
Little evidence of a shape difference is exhibited. Confidence is gained that the full-geometry analysis can be used to improve the statistics, particularly at the forward angles.
The same comparison is made for the full geometry, 20-MeV threshold analysis in the bottom half of
Fig.~\ref{fig:cxs_comp_298}.  The average of these yields is 0.45\% higher than the standard
analysis.  The reproducibility of the acceptance-corrected yields at the 1\% level for conditions in which the acceptance changes by as much as 30\% lends credibility to the Monte Carlo simulation.

\subsection{Beam momentum}

The momentum calibration of the C6 and C8 beam lines have been checked extensively in
previous experiments, including two recent publications from our
collaboration\cite{Starostin01, Tippens01}.
The good energy and spatial resolution of the Crystal Ball can be utilized to determine the 
pion beam momentum at target center.  The procedure was as follows:

\begin{itemize}

\item{The overall gain of the NaI crystals was adjusted so that the centroid of the
invariant mass spectrum of two-cluster events equaled the $\pi^0$ mass. A similar
procedure was applied to the Monte Carlo simulation.}
\item{The data were analyzed 
assuming different values of the ``real" beam momentum.  Monte Carlo
events were generated and analyzed at the same intervals of the beam momentum (1 MeV/$c$).  The Monte Carlo events were distributed in angle as predicted by the recent GW SAID FA02 analysis\cite{said} at the nominal momentum.}
\item{The difference in the missing mass was plotted as a function of
the momentum and found to be linear.  A linear fit of the missing mass difference
was performed.}
\item{The solution of the linear fit where the difference was zero was chosen as the
correct central beam momentum at target center.}

\end{itemize}

This technique gives the average momentum of the pions that produced charge-exchange
events. These results can be compared to the pion momentum at target center by
subtracting from the calibrated momenta the momentum loss in the beam scintillators, air,
vacuum windows and half of the length of the LH$_2$ target.  Table~\ref{tab:momtab} shows
the values of the pion momenta obtained from these methods.

\begin{table}
\caption{\label{tab:momtab}Comparison of the beam momenta (in MeV/$c$) from CB analysis and the
C6 channel calibration corrected for the momentum loss between the channel and target center.}
%\begin{ruledtabular}
\begin{tabular}{|c|c|}
\hline
\hline
 CB Analysis&C6 Calibration\\
\hline
147.9&146.6\\
173.8&174.0\\
188.3&186.7\\
212.3&209.8\\
237.9&236.3\\
271.2&268.0\\
298.3&296.5\\
322.8&321.3\\
\hline
\hline
\end{tabular}
%\end{ruledtabular}
\end{table}

The momenta from the CB analysis in Table~\ref{tab:momtab} are used for the present
results. The first dipole in the C6 line was adjusted slightly as part of the beam tuning
procedure in order to center the beam on the target, which can produce small deviations
from the nominal momentum for a given tune.  The momenta from the C6 calibration are
systematically lower by an average amount of 1.7 MeV/$c$, which is adopted as the estimated
uncertainty in the momenta.

\begin{figure*}
 \resizebox{0.90\textwidth}{!}{\includegraphics{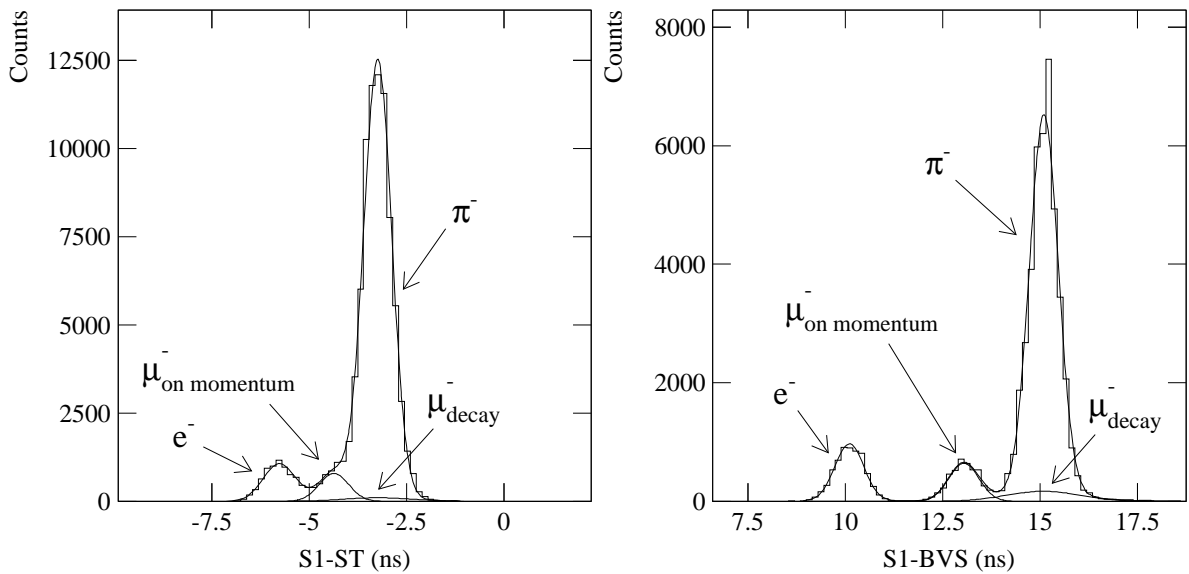}}
\caption{TOF spectra at 238 MeV/$c$ for S1-ST (left) and S1-BVS (right).
The electrons are the left-most peak in both spectra, followed by on-momentum muons and pions.
Decay muons fall under the pion peak.
The fit consisting of four gaussians is shown as well as the individual contributions of the
on-momentum muons and decay muons.  The fitting procedure is discussed in the text.}
 \label{fig:tof_238}
\end{figure*}

\subsection{Beam contamination and systematics}

The contamination of muons and electrons in the beam were evaluated using
\mbox{time-of-flight} (TOF). This technique limited the upper momentum to 323 MeV/$c$ in
order to provide adequate separation between $\pi$'s, $\mu$'s and $e$'s.  Data at higher
momenta (up to 750 MeV/$c$) require a separate analysis of the electron contamination from the Cherenkov counter and will be published at a later time. Pion fractions at ST were determined directly
from the S1-ST TOF (5.2 m flight path, see Fig.~\ref{fig:cryball}b) at the four lowest
momenta.  The S1-BVS TOF (8.9 m flight path) was used at the four highest momenta, which
required a correction back to ST.  

Sample TOF spectra are shown at 238 MeV/$c$ in Fig.~\ref{fig:tof_238} for both S1-ST and
S1-BVS. The S1-BVS TOF was used to determine the contaminations at this and higher momenta
due to the overlap of the small muon peak with the pion peak in the S1-ST spectrum. 

The on-momentum
muons in the middle peak originate from pions that decay in the vicinity of the
production target and fall into the acceptance of the beam channel.  Muons that originate
from pion decay in the beam line before the last magnetic element typically fall outside
of the channel acceptance. Muons that originate from pion decay after the last beam channel
magnet cannot be distinguished from pions in the TOF.
A correction to the pion area was made for these so-called decay muons that hit either ST or BVS. 
This fraction was determined from a beam line Monte Carlo program based on GEANT
\cite{geant}.  The simulation started at the exit of the last quadrupole with the
trajectories that were determined from the beam drift chambers. The ratio of decay muons
to pions ranged from 1.8\% (6.2\%) at ST (BVS) at 323 MeV/$c$ to 2.7\% (8.8\%) at 148
MeV/$c$.  The BVS percentages were higher due to its larger size (15.2 cm x 15.2 cm x 0.6
cm thick) and the decay of pions between ST and BVS.

The simulated TOF distributions for the decay muons peaked near the pion peak but had
tails on both sides corresponding to forward- and backward-going muons in the pion frame.
Thus four gaussian peaks were fitted to determine the peak areas corresponding to $e$'s, on-momentum $\mu$'s,  $\pi$'s and decay $\mu$'s.  The ratio of decay $\mu$'s to $\pi$'s in the fit was forced to be that predicted by the beam Monte Carlo simulation.  

Other constraints were utilized in the fits.  Widths of the peaks were averaged for the different particles
at lower momenta where the peaks were well separated and applied as constraints at the higher
momenta where they overlapped.  The positions were constrained by calculating the positions
for the different momenta and applying a small linear correction determined empirically
at the lower momenta. The on-momentum muon areas in the fits to the S1-ST TOF at the four
highest momenta were constrained by assuming that the ratio of these muons that appeared in
the S1-BVS TOF was the same as for electrons.  This assumption was verified by the beam line
Monte Carlo simulation and from the analyses at the lower momenta.

\begin{figure*}
 \resizebox{0.9\textwidth}{!}{\includegraphics{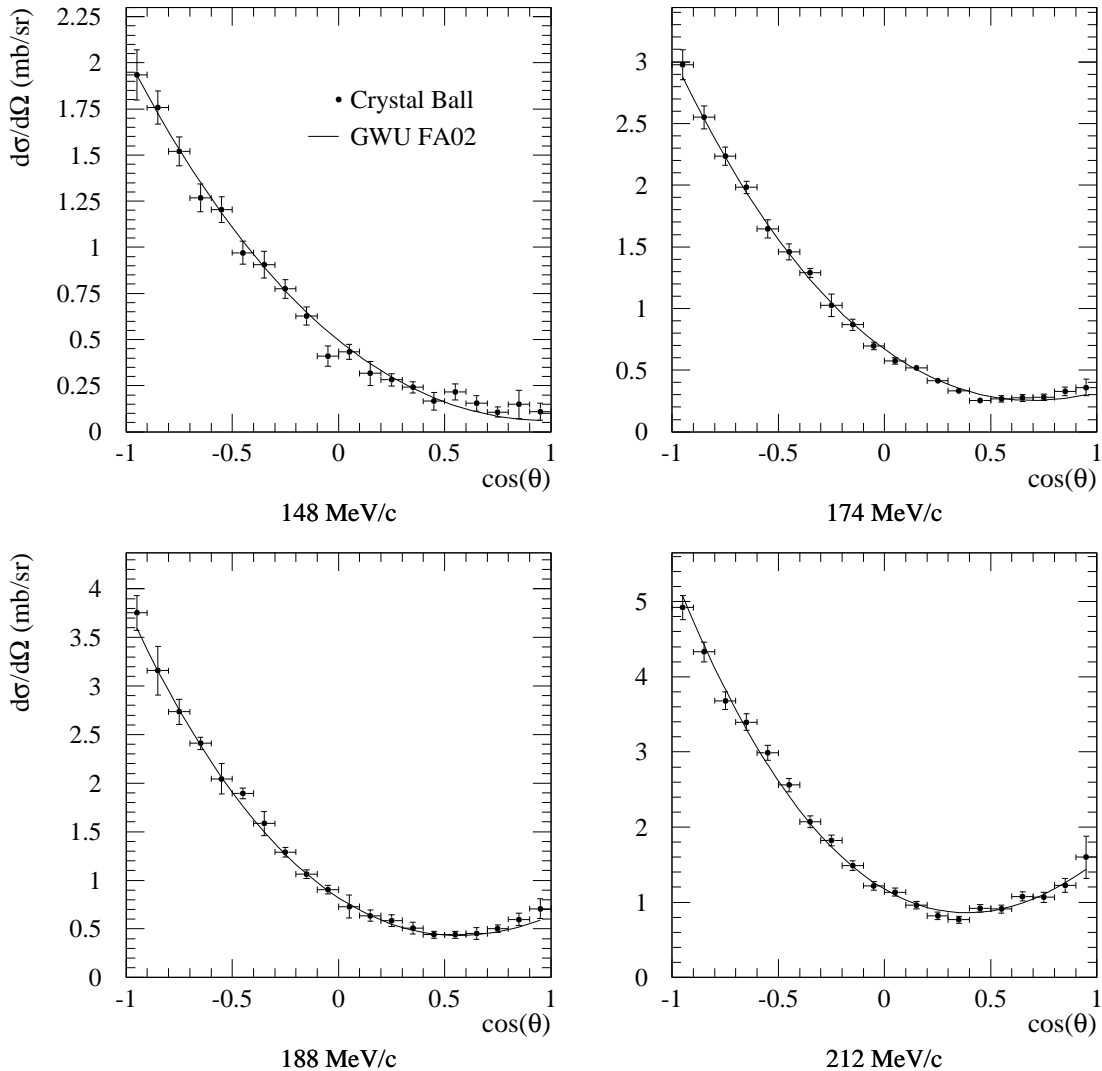}}
 \caption{Differential cross sections of reaction $\pi^- p \to \pi^0 n$. Black circles are the values obtained
  in this experiment. The curves show the results of the FA02 partial-wave analysis of the George Washington group \cite{said}
  based on experiments made earlier by other groups. }
 \label{fig:dcs1}
\end{figure*}

\begin{figure*}
 \resizebox{0.9\textwidth}{!}{\includegraphics{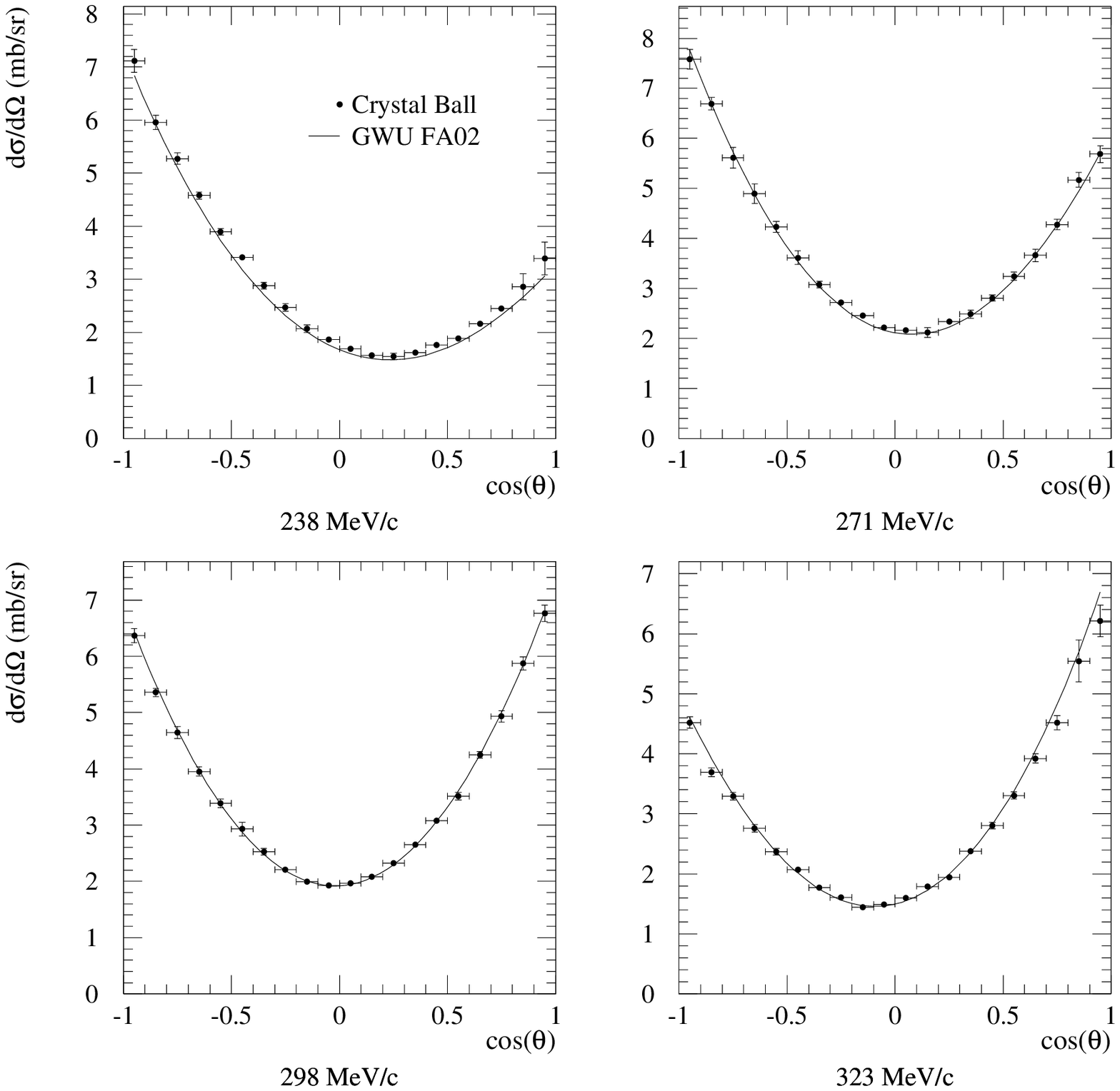}}
 \caption{Differential cross sections of reaction $\pi^- p \to \pi^0 n$. Black circles are the values obtained
  in this experiment. The curves show the results of the FA02 partial-wave analysis of the George Washington group \cite{said}
  based on experiments made earlier by other groups.}
 \label{fig:dcs2}
\end{figure*}

\begin{figure*}
 \resizebox{0.9\textwidth}{!}{\includegraphics{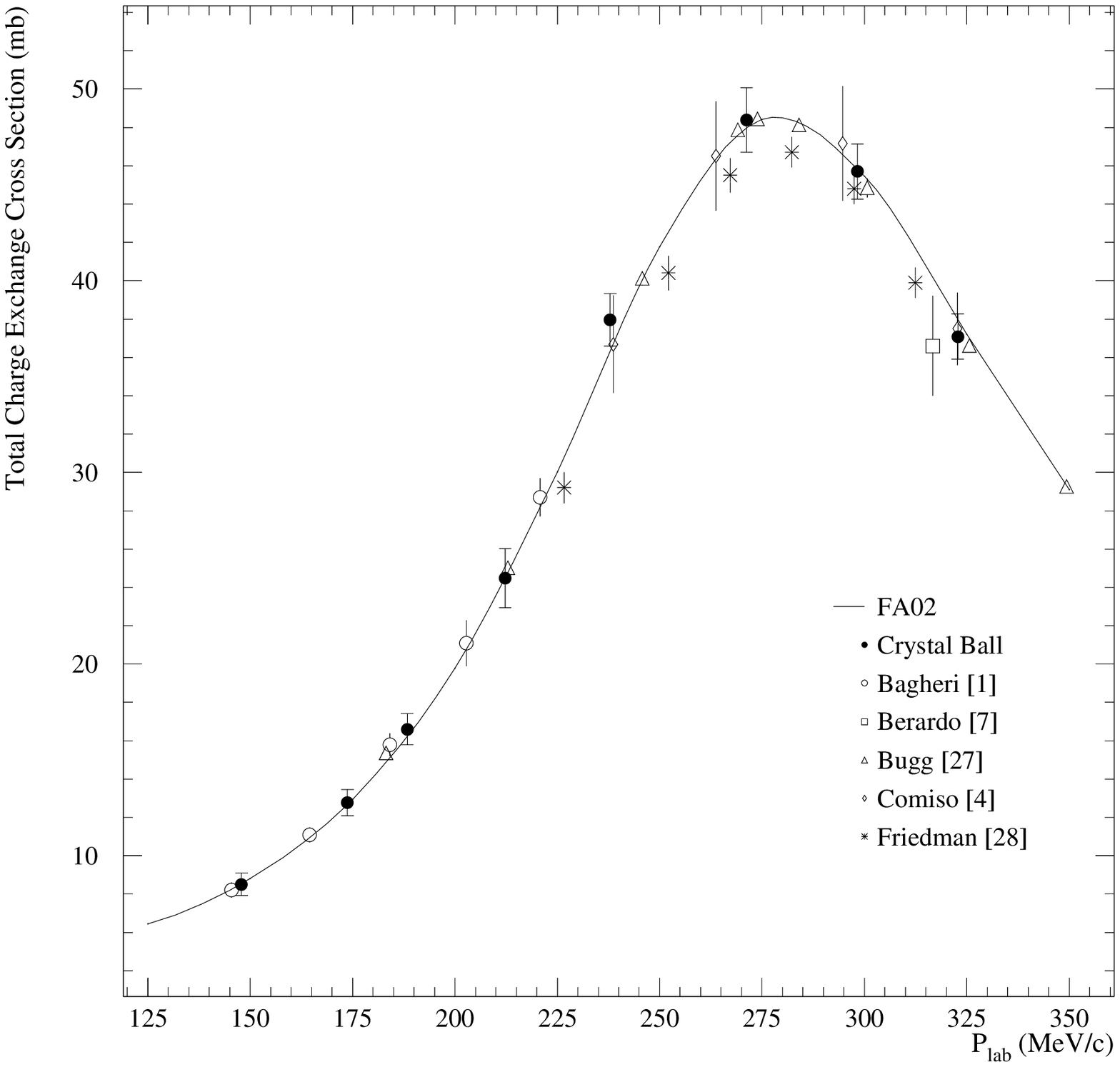}}
\caption{The total charge-exchange cross section obtained from integrating the
differential cross section.  The error bars show combined statistical and systematic uncertainties as described in the text.  The results are compared
to the GWU FA02 partial-wave analysis\cite{said}
and to previous data\cite{lit1,lit4,lit7,Bugg71,Friedman93}}.
 \label{fig:totcxs}
\end{figure*}

Corrections were applied for decay and multiple scattering of beam pions between ST and the target.
These corrections were determined using the beam line Monte Carlo program to start with pions at the center
of the drift chambers and propagate them to the target center.
The fraction of pions within the target radius at target center to the number traversing ST was recorded.
The multiple scattering losses are significant, resulting in an additional reduction of pions at target center
(compared to decay alone) of 5\% at 323 MeV/c and 19\% at 148 MeV/c.  An uncertainty of 20\% of this correction was applied and is the dominant contributor to the overall systematic uncertainty at the lower momenta.

\section{Results}

The obtained values of $\pi^- p \to \pi^0 n$ differential cross sections are shown in
Tables~\ref{tab:dcs1}~and~\ref{tab:dcs2}.  They are plotted in Fig.~\ref{fig:dcs1} and
Fig.~\ref{fig:dcs2} together with the results of the FA02 partial-wave analysis of the
George Washington group \cite{said}. The statistical uncertainties of the differential cross
section are typically 2-6\% except at the lowest momentum and the forward-angle points at the three lowest momenta where the cross sections decrease to a few tenths of a mb.

A minimum systematic uncertianty of 2.0\% was applied at all momenta to account for the calibration of the veto barrel, the uncertainty of determining the probability of vetoing legitimate events in the
veto scintillators.  An additional 1.5\% was added at all momenta to account for the uncertainties in effective target length, hydrogen density, and the residual gas in the target for the empty runs.  The following systematic uncertainties were included in Tables~\ref{tab:dcs1}~and~\ref{tab:dcs2}:
1) the uncertainties in the fits of the pion peak in the TOF spectra ($\approx$1\%),  2) the statistical
uncertainty for the counts in the pion peak in the TOF spectra (0.5 - 1.4\%), and 3) 20\% of the multiple scattering losses to the pion beam (1.1 - 5.9\%).  The quadrature summation of these factors gives total
systematic uncertainties of 3.1\% to 6.5\%, increasing as the beam momentum
decreases.

The data presented here were analyzed independently at ACU and PNPI.  Consensus was obtained on the systematic factors and initial differences in the separate analyses were useful in estimating the systematic uncertanties.  Independent energy calibrations, cuts, acceptance calculations and the like produced point-to-point differences in the results between the analyses. These differences were almost always smaller than the statistical uncertainties, in which case the cross section reported is the weighted average and the uncertainty is the simple average. For the cases where the cross section differed by more than the statistical uncertainty the uncertainties were increased so that they extended to the points obtained from the separate analyses.

The differential cross sections were integrated to obtain the total charge-exchange cross sections at the eight momenta. These cross sections, statistical uncertainties and
total uncertainties are listed in Table~\ref{tab:totcxs}.
The systematic uncertainty was added in quadrature to the statistical uncertainty for the total uncertainty.  The results are shown in Fig.~\ref{fig:totcxs}.  As with the differential cross sections, the general agreement with the GWU FA02 partial-wave analysis is good.  The most accurate data on which the partial-wave analysis is based are
Ref.\cite{Bugg71,Friedman93}.  These experiments measured the fraction of beam pions that converted to neutral final states in a hydrogen target and made corrections for small effects such as $\pi^- p \to \gamma n$.

\section{Conclusion}

Differential cross sections of the charge-exchange reaction $\pi^- p \to \pi^0 n$ are presented in the region of the $\Delta$(1232) resonance.  The present
results nearly double the database for these measurements in this momentum
interval. Complete angular coverage is provided at all momenta using the Crystal
Ball multi-photon spectrometer.

The obtained cross sections are in good agreement with the results of the GWU FA02 partial-wave analysis based on earlier experiments.  These data provide more robust input for determinations of the mass and width splitting of the
$\Delta^0$ - $\Delta^{++}$ resonances and to investigate isospin breaking using
partial-wave analyses, potential models or chiral Lagrangians. The data will be useful for obtaining some of the most important numbers in hadronic physics: the $\pi$NN coupling constant, the $\pi$N sigma term, and the up-down quark mass difference.
  
%\begin{thebibliography}{99}
% \bibitem{slac}  E.D.~Bloom and C.W.~Peck, Ann. Rev. Nucl. Sci. 33, 143 (1983).
% \bibitem{geant} GEANT 3.21 CERN Program Library Long Writeup W5013, CERN, Geneva, Switzerland.
% \bibitem{said}  SAID at $http://GWDAC.phys.GWU.EDU$
%\end{thebibliography}

%\begin{table*}
%\caption{\label{tab:table3}This is a wide table that spans the page
%width in \texttt{twocolumn} mode. It is formatted using the
%\texttt{table*} environment. It also demonstates the use of
%\textbackslash\texttt{multicolumn} in rows with entries that span
%more than one column.}
%\begin{ruledtabular}
%\begin{tabular}{ccccc}
% &\multicolumn{2}{c}{$D_{4h}^1$}&\multicolumn{2}{c}{$D_{4h}^5$}\\
% Ion&1st alternative&2nd alternative&lst alternative
%&2nd alternative\\ \hline
% K&$(2e)+(2f)$&$(4i)$ &$(2c)+(2d)$&$(4f)$ \\
% Mn&$(2g)$\footnote{The $z$ parameter of these positions is $z\sim\frac{1}{4}$.}
% &$(a)+(b)+(c)+(d)$&$(4e)$&$(2a)+(2b)$\\
% Cl&$(a)+(b)+(c)+(d)$&$(2g)$\footnotemark[1]
% &$(4e)^{\text{a}}$\\
% He&$(8r)^{\text{a}}$&$(4j)^{\text{a}}$&$(4g)^{\text{a}}$\\
% Ag& &$(4k)^{\text{a}}$& &$(4h)^{\text{a}}$\\
%\end{tabular}
%\end{ruledtabular}
%\end{table*}

\begin{table*}
\caption{\label{tab:dcs1}Differential cross sections [mb/sr] and statistical uncertainties for the reaction $\pi^- p \to \pi^0 n$.  The systematic uncertainty at each momentum is given as a percentage.}
\begin{ruledtabular}
\begin{tabular}{|c|cc|cc|cc|cc|}
Momentum&\multicolumn{2}{c|}{148 MeV/c}&\multicolumn{2}{c|}{174 MeV/c}&\multicolumn{2}{c|}{188
MeV/c}&\multicolumn{2}{c|}{212 MeV/c}\\
\hline
Systematic unc.&\multicolumn{2}{c|}{6.5\%}&\multicolumn{2}{c|}{5.2\%}&\multicolumn{2}{c|}{4.5\%}&\multicolumn{2}{c|}{4.0\%}\\
\hline
 $\quad\cos\theta_{cm}\quad$&$\quad d\sigma/d\Omega\quad$&\quad unc.\quad&$\quad d\sigma/d\Omega\quad$&\quad unc.\quad&
                             $\quad d\sigma/d\Omega\quad$&\quad unc.\quad&$\quad d\sigma/d\Omega\quad$&\quad unc.\quad\\
\hline
-0.95&1.934&0.136&2.976&0.121&3.752&0.180&4.920&0.161\\
-0.85&1.757&0.089&2.550&0.094&3.159&0.251&4.330&0.132\\
-0.75&1.520&0.079&2.234&0.073&2.734&0.130&3.681&0.115\\
-0.65&1.267&0.075&1.983&0.050&2.409&0.066&3.397&0.108\\
-0.55&1.204&0.069&1.646&0.073&2.045&0.158&2.987&0.099\\
-0.45&0.970&0.062&1.459&0.065&1.894&0.054&2.561&0.089\\
-0.35&0.905&0.073&1.288&0.038&1.584&0.126&2.071&0.078\\
-0.25&0.774&0.052&1.025&0.092&1.291&0.049&1.822&0.072\\
-0.15&0.627&0.050&0.868&0.045&1.064&0.042&1.487&0.064\\
-0.05&0.410&0.054&0.696&0.029&0.903&0.042&1.219&0.057\\
 0.05&0.434&0.040&0.575&0.027&0.728&0.118&1.135&0.055\\
 0.15&0.317&0.065&0.519&0.024&0.636&0.059&0.962&0.051\\
 0.25&0.282&0.033&0.414&0.024&0.584&0.060&0.820&0.047\\
 0.35&0.241&0.031&0.332&0.022&0.506&0.061&0.768&0.046\\
 0.45&0.166&0.047&0.253&0.024&0.439&0.036&0.919&0.052\\
 0.55&0.217&0.043&0.266&0.027&0.439&0.037&0.909&0.053\\
 0.65&0.154&0.043&0.274&0.029&0.452&0.059&1.077&0.062\\
 0.75&0.107&0.028&0.280&0.026&0.501&0.039&1.066&0.068\\
 0.85&0.149&0.077&0.326&0.035&0.597&0.063&1.224&0.091\\
 0.95&0.110&0.046&0.360&0.069&0.707&0.103&1.600&0.282\\  
\end{tabular}
\end{ruledtabular}
\end{table*}

\begin{table*}
\caption{\label{tab:dcs2}Differential cross sections [mb/sr] for the reaction $\pi^- p \to \pi^0 n$.}
\begin{ruledtabular}
\begin{tabular}{|c|cc|cc|cc|cc|}
Momentum&\multicolumn{2}{c|}{238 MeV/c}&\multicolumn{2}{c|}{271 MeV/c}&\multicolumn{2}{c|}{298
MeV/c}&\multicolumn{2}{c|}{323 MeV/c}\\
\hline
Systematic unc.&\multicolumn{2}{c|}{3.5\%}&\multicolumn{2}{c|}{3.4\%}&\multicolumn{2}{c|}{3.1\%}&\multicolumn{2}{c|}{3.1\%}\\
\hline
 $\quad\cos\theta_{cm}\quad$&$\quad d\sigma/d\Omega\quad$&\quad unc.\quad&$\quad d\sigma/d\Omega\quad$&\quad unc.\quad&
                             $\quad d\sigma/d\Omega\quad$&\quad unc.\quad&$\quad d\sigma/d\Omega\quad$&\quad unc.\quad\\
\hline

-0.95&7.117&0.214&7.579&0.196&6.370&0.127&4.521&0.098\\
-0.85&5.956&0.134&6.694&0.120&5.360&0.073&3.694&0.070\\
-0.75&5.273&0.107&5.610&0.206&4.646&0.106&3.290&0.063\\
-0.65&4.582&0.066&4.889&0.196&3.953&0.083&2.756&0.063\\
-0.55&3.897&0.061&4.230&0.109&3.387&0.080&2.370&0.052\\
-0.45&3.408&0.055&3.614&0.134&2.929&0.122&2.070&0.048\\
-0.35&2.880&0.065&3.073&0.067&2.527&0.062&1.769&0.042\\
-0.25&2.468&0.070&2.713&0.061&2.207&0.051&1.606&0.040\\
-0.15&2.071&0.074&2.459&0.057&1.994&0.033&1.443&0.036\\
-0.05&1.862&0.046&2.215&0.053&1.927&0.032&1.491&0.037\\
 0.05&1.690&0.040&2.166&0.052&1.966&0.032&1.597&0.038\\
 0.15&1.570&0.035&2.117&0.098&2.081&0.033&1.790&0.040\\
 0.25&1.545&0.065&2.335&0.055&2.325&0.036&1.946&0.042\\
 0.35&1.614&0.036&2.484&0.082&2.651&0.040&2.374&0.048\\
 0.45&1.758&0.038&2.808&0.063&3.077&0.043&2.803&0.054\\
 0.55&1.881&0.051&3.242&0.084&3.513&0.066&3.307&0.062\\
 0.65&2.158&0.058&3.658&0.120&4.248&0.058&3.924&0.075\\
 0.75&2.447&0.056&4.277&0.100&4.934&0.103&4.517&0.115\\
 0.85&2.861&0.247&5.170&0.149&5.872&0.118&5.547&0.352\\
 0.95&3.393&0.308&5.686&0.169&6.765&0.147&6.217&0.262\\
\end{tabular}
\end{ruledtabular}
\end{table*}

\begin{table*}
\caption{\label{tab:totcxs}Total charge-exchange reaction cross sections derived from integrating the differential cross sections.  Statistical and total uncertainties are included.}
%\begin{ruledtabular}
\begin{tabular}{|c|c|c|c|}
\hline
\hline
Momentum&Total Cross Section (mb)&Statistical Uncer.&Total Uncer.\\
\hline
147.9&8.5&0.3&0.9\\
173.8&12.8&0.3&1.1\\
188.3&16.6&0.5&1.2\\
212.3&24.5&0.5&2.4\\
237.9&38.0&0.5&2.1\\
271.2&48.4&0.5&2.6\\
298.3&45.7&0.4&2.3\\
322.8&37.1&0.5&1.9\\
\hline
\hline
\end{tabular}
%\end{ruledtabular}
\end{table*}

\begin{acknowledgments}
% put your acknowledgments here.
This work was supported in part by the U.S. DOE and NSF, by the Russian Foundation for Basic Research, and
by NSERC of Canada.
\end{acknowledgments}

% Create the reference section using BibTeX:
\bibliography{prccex}

\end{document}